\documentclass[preprint,showpacs,preprintnumbers,amsmath,amssymb]{revtex4}

\usepackage{graphicx}% Include figure files
\usepackage{dcolumn}% Align table columns on decimal point
\usepackage{bm}% bold math
\begin{document}
%\preprint{}
\title{Tunnelling of polarized electrons in magnetic wires
}% Force line breaks with \\
\author{D. Schmeltzer}

\affiliation{Physics Department \\ City College of the City
University of New York
\\ New York, New York 10031}%Lines break automatically or can be forced with \\

\date{\today}% It is always \today, today,
\begin{abstract}
We investigate the tunnelling between an electronic gas with two
different velocities $V_\uparrow \neq V_\downarrow$ and a regular
metal.  We find that at the interface between the two systems that
the tunnelling conductance for spin up and spin down are different.
As a result a partly polarized gas
(``magnetic wire'') is obtained.  This result is caused by the e-e
interaction, $g_2'' - g_1'' \neq 0$, which in the presence of
$V_\uparrow \neq V_\downarrow$ gives rise to two different
tunnelling exponents.
\end{abstract}

\maketitle
%----------------------------------------------------------------------
Following ideas of transport in ``Quantum Wires'' we can study
transport and tunnelling of polarized electrons.  Studies of
polarized electrons in the diffusive regime have been considered
by Aronov [1].  Recently the spin analog of the bipolar p-n-p
transistor and F.E.T. have been suggested [2-4].  These studies
have inspired us to study tunnelling in the ``Quantum Wire''
regime between a ferromagnetic wire and a metallic one.  The
analog of the p-n-p transistor can be simulated with the help of
magnetic wires using two ferromagnetic metallic (FM) wires coupled
to a metallic (M) dot.  We obtain a double junction FM-M-FM.  When
the ferromagnetic wires have parallel polarization, we obtain a
low resistance device which is switched of when the two wires are
antiparallel polarized.

As a first step in this direction we will investigate the
tunnelling between a quantum wire in the presence of interaction
with two different Fermi velocities (for spin up and spin down)
to a metallic wire.  The origin of the two Fermi velocities is the
presence of an external magnetic field
and/or the presence of Ferromagnetic interactions.  As a result,
we have a situation where the two channels (spin up and spin down)
have two different velocities, $V_\uparrow$, $V_\downarrow$.  In
one dimension the ballistic conductance is independent from the
velocity.  Therefore in the absence of electron electron
interaction the ballistic conductance for spin up and down are
equal $G_\uparrow = G_\downarrow = e^2/h$ despite of the fact that
the two Fermi velocities are different, $V_\uparrow \neq
V_\downarrow$.  In order to obtain $G_\uparrow \neq G_\downarrow$
we have to take in consideration the electron-electron
interaction.  The electron-electron interaction in one dimension
causes fractionalization of the electron.  This is observed in
tunnelling experiments where the interaction gives rise to an
orthogonality catastrophe controlled by $K = V_F / V_P$ (``$V_F$''
is the Fermi velocity and ``$V_P$'' is the velocity of the
collective excitations)[5].  For a polarized gas $V_\uparrow >
V_\downarrow$ one has two interaction parameters $K_\uparrow =
V_\uparrow / V_P$, $K_\downarrow = V_\downarrow / V_P$.  Due to
the fact that $V_\uparrow > V_\downarrow$ we obtain $1 >
K_\uparrow > K_\downarrow$. As a result the tunnelling for the two
channels will be different.  The tunnelling conductance between a
polarized gas with interaction and a metallic one is suggested to
be: $G_\uparrow \simeq \frac{e^2}{h} |t|^2
\ell^{2(\frac{1}{K_\uparrow} -1)}$, $G_\downarrow \simeq
\frac{e^2}{h} |t|^2 \ell^{2(\frac{1}{K_\downarrow} -1)}$ where
``t'' is the tunnelling matrix between the magnet wire and the
metallic one and $\ell = \frac{1}{a} \min[L, L_T]$, ``$a$'' is the
lattice spacing, $L$ is the length of the wire and $L_T =
\frac{\hbar V_F}{k_B T}$ is the thermal length.  In the limit of
long wires $L> L_T$,  the tunnelling is controlled by the
temperature and one can replace ``$\ell$'' with $\frac{T_F}{T}$.
$\tilde{T}_F$ is the thermal ultraviolet cutoff, $\tilde{T}_F =
\frac{\hbar V}{K_B a} $, replacing $1/a$ with the Fermi momentum
$K_F$ and the velocity $V$ with $V_F$ we obtain the Fermi
temperature, $T_F = \frac{\hbar V_F K_F}{K_B}$. (In our problem we
have two pairs of Fermi velocities, $V_\uparrow$, $V_F$ and
$V_\downarrow$, $V_F$, the use of Fermi temperature $T_F$ as a U.V.
cutoff is an approximation).  The polarization degree will be
controlled by the difference $ \frac{1}{K_\uparrow} -
\frac{1}{K_\downarrow} $.  Therefore, increasing the difference
$V_\uparrow - V_\downarrow$, lowering the temperature and
increasing the electron-electron interaction will enhance the
anisotropy of the conductance $\frac{G_\downarrow}{G_\uparrow} =
(\frac{T}{T_F})^{2 \frac{K_\uparrow - K_\downarrow}{K_\uparrow
K_\downarrow}}$. Since $K_\uparrow < K_\downarrow $, we obtain at
low temperature a strong polarization, $
\frac{G_\downarrow}{G_\uparrow} \rightarrow 0$. When $T
\rightarrow 0$ the value of the conductance is controlled by the
length of the wire ``L'' since $L < L_T$, we obtain
$\frac{G_\downarrow}{G_\uparrow} = (\frac{a}{L})^{2
\frac{K_\uparrow - K_\downarrow}{K_\uparrow K_\downarrow}}$.

We
find that the interaction parameter $g_2'' - g_1''$ (the
difference between the forward and backward interaction) controls
the polarization.  In particular when $g_1 < 0$ (attractive
interaction with large momentum transfer), the polarization is
enhanced.

Next we present the model and the calculation.  The hamiltonian it
is given by $H^{(1)} + H^{(2)} + H_T$.  $H^{(1)}$ represents a
polarized channel with electron-electron interaction.  $H^{(2)}$
is a metallic wire and $H_T$ is the tunnelling part between the
two wires.

The first wire is restricted to the region $ - L \leq x \leq0$.
$H_0^{(1)}$ is the kinetic energy with two velocities
$V_\uparrow$,$V_\downarrow$,$V_\uparrow - V_\downarrow = \Delta$.
$H_I^{(1)}$ is the e-e interaction away from half-filling,
\begin{widetext}
\begin{eqnarray}\nonumber
H_I^{(1)} = \int_{-L}^0 dx \sum_{\sigma, \sigma'} \{ g_1
R_\sigma^+(x) L_{\sigma'}^+(x) R_{\sigma'}(x) L_\sigma(x)
+ g_2 R_\sigma^+(x) L_{\sigma'}^+(x) L_{\sigma'}(x) R_\sigma(x) \\
 + g_4 (1-\delta_{\sigma, \sigma'}) [R_\sigma^+(x)
R_{\sigma'}^+(x) R_{\sigma'}(x) R_\sigma(x) + \; R \leftrightarrow
L \; ] \; \}   \qquad
\end{eqnarray}
\end{widetext}
The form of the interaction hamiltonian given in eq.(1) has been
obtained with the help of the fermion operators
$\tilde{\psi}_\sigma(x)$,$\tilde{\psi}_\sigma^+(x)$, $\sigma =
\uparrow, \downarrow$ in terms of the right and left movers,
$\tilde{\psi}_\sigma(x) = e^{iK_{F,\sigma}x} R_\sigma(x) +
e^{-iK_{F,\sigma}x} L_\sigma(x)$. $g_1 = g_1^\perp(0) = g_1''(0)$
denotes the backward $(K_F, -K_F) \rightarrow (-K_F, K_F)$ and
$g_2 = g_2^\perp(0) = g_2''(0)$ denote the forward $(-K_F, K_F)
\rightarrow (-K_F, K_F)$ scattering potentials.

The backward interaction can be split into $g''$ and $g_1^\perp$,
$g_1''$ controls the spin and charge density interaction.  This
notation emphasizes the connection with the Kondo parameters -
$J_\parallel$ and $J_\perp$.

We will bosonize the hamiltonian $H^{(1)}$ using open boundary
conditions, $\tilde{\psi}_\sigma(0)= \tilde{\psi}_\sigma(-L) = 0$.
We will follow reference [6] and we will introduce a chiral
fermion $\psi_\sigma(x)$ which will obey periodic boundary
condition and can be bosonized according to the standard rules.
\begin{subequations} \label{eq2:whole}
\begin{equation}
\psi_\sigma(x) = \left\{
\begin{array}{c}
  R_\sigma(x)  \quad x > 0 \\
  L_\sigma(-x) \quad x < 0
\end{array} \right.
,\quad R_\sigma(x) = - L_\sigma(-x) \label{subeq2:1}
\end{equation}
\begin{equation}
\psi_\sigma(x) = \frac{1}{\sqrt{2\pi d}} \exp[i \sqrt{4 \pi}
\theta_\sigma(x)] \label{subeq2:2}
\end{equation}
\begin{equation}
J_\sigma(x) = : \psi_\sigma^+(x) \psi_\sigma(-x): =
\frac{1}{\sqrt{\pi}} \partial_x \theta_\sigma(x)  \label{subeq2:3}
\end{equation}
\begin{equation}
J_\sigma(x) = \left\{
\begin{array}{c}
  : R_\sigma^+(x) R_\sigma(x):  \quad x > 0 \\
  : L_\sigma^+(x) L_\sigma(-x):  \quad x < 0
\end{array} \right.  \label{subeq2:4}
\end{equation}
\end{subequations}
$J_\sigma(x)$ and $\theta_\sigma(x)$ are the chiral current and
chiral bosonic field.  Using the chiral boson representation we
obtain the bosonic representation for the region $-L < x \leq 0$.
\begin{widetext}
\begin{subequations} \label{eq3:whole}
\begin{equation} \nonumber
H^{(1)} = H_0^{(1)} + H_I^{(1)}
\end{equation}
\begin{equation}
H_0^{(1)} = \int_{-L}^L dx \{\pi V_\uparrow J_\uparrow^2(x) + \pi
V_\downarrow J_\downarrow^2(x) \}, \quad V_\uparrow - V_\downarrow
= \Delta \label{subeq3:1}
\end{equation}
\begin{eqnarray} \nonumber
H_I^{(1)} &=& \int_{-L}^L dx \{ \frac{1}{2} ( g_2^\parallel -
g_1^\parallel)(J_\uparrow(x) J_\uparrow(-x) + J_\downarrow(x)
J_\downarrow (-x) \\ \nonumber &+& \frac{1}{2} g_2^\perp (
J_\uparrow(x) J_\downarrow(-x) + J_\downarrow(x) J_\uparrow(-x)) )
+ g_4 J_\uparrow(x) J_\downarrow(x) \\ &+& \frac{g_1^\perp}{2(\pi
d)^2} \cos[ \sqrt{4\pi}(\theta_\uparrow(x) +
\theta_\uparrow(-x)-\theta_\downarrow(x) - \theta_\downarrow(-x))+
2 K_F \left(\frac{\Delta}{2 V_F}\right)x \,] \; \}
\label{subeq3:2}
\end{eqnarray}
\end{subequations}
\end{widetext}
From eq.3b we observe that the Luttinger parameter
which controls separately the spin up and spin down excitations is
given by $g_2'' - g_1''$.  The difference in the velocities
$V_\uparrow - V_\downarrow = \Delta$ represents the Zeeman energy and
$\frac{V_\uparrow + V_\downarrow}{2} = V_F$ is the Fermi velocity.
The last term in eq. 3b will be ignored here since it contains the
oscillating term $2 K_F (\frac{\Delta}{2 V_F} ) x $.

We will investigate the case with the Zeeman term $\Delta \neq 0$
and therefore we will neglect the cosine term.  We will
diagonalize the hamiltonian in eqs. 3a - 3b performing a set of
transformations which leaves the commutation rules invariant.

We introduce two new chiral currents $J_+(x) $ and $J_-(x)$
\begin{subequations} \label{eq4:whole}
\begin{eqnarray} \nonumber
J_\uparrow (x) = a J_+(x) + b J_-(x) \\ \nonumber J_\downarrow(x)
= b J_+(x) - a J_-(x) \label{subeq4:1}
\end{eqnarray}
where $a^2 + b^2 = 1$, and a second transformation for the new
variables $\tilde{J}_+(x)$,$\tilde{J}_-(x)$.
\begin{equation}
J_\pm(x) = \frac{K_\pm^{1/2}}{2} (\tilde{J}_\pm(x) +
\tilde{J}_\pm(-x) ) + \frac{K_\pm^{-1/2}}{2} (\tilde{J}_\pm(x) -
\tilde{J}_\pm(-x) ) \label{subeq4:2}
\end{equation}
\end{subequations}
The set of parameters $K_+$ and $K_-$ are controlled by the
interaction term $\frac{g_2'' - g_1''}{2 \pi V_\pm}$.  As a result
the hamiltonian in eq. 3 takes the form:
\begin{widetext}
\begin{subequations} \label{eq5:whole}
\begin{eqnarray}
H^{(1)} &=& \int_{-L}^L dx \Bigg\{ \pi \tilde{V}_+
\tilde{J}_+^2(x) + \pi \tilde{V}_- \tilde{J}_-^2(x) \\ \nonumber
&+& \Bigg[ g_2^\perp (b^2 - a^2)\frac{1}{2}\bigg( \big(
\frac{K_+}{K_-}\big)^{\frac{1}{2}} +
\big(\frac{K_+}{K_-}\big)^{-\frac{1}{2}} \bigg) \\ \nonumber &&+
\Big(g_4(b^2 - a^2) + 2 \pi a b (V_\uparrow -
V_\downarrow)\Big)\frac{1}{2}
\bigg(\frac{K_+}{K_-}\bigg)^{\frac{1}{2}} -
\bigg(\frac{K_+}{K_-}\bigg)^{-\frac{1}{2}} \Bigg] \tilde{J}_+(x)
\tilde{J}_-(-x) \\ \nonumber &+& \Bigg[ g_2^\perp (b^2 - a^2)
\frac{1}{2} \bigg( \big(\frac{K_+}{K_-}\big)^{\frac{1}{2}} - \big(
\frac{K_+}{K_-} \big)^{-\frac{1}{2}} \bigg) \\ \nonumber &&+
\Big(g_4 (b^2 - a^2) + 2 \pi a b (V_\uparrow - V_\downarrow) \Big)
\frac{1}{2}\bigg(\frac{K_+}{K_-}\bigg)^{\frac{1}{2}} +
\bigg(\frac{K_+}{K_-}\bigg)^{-\frac{1}{2}} \Bigg] \tilde{J}_+(x)
\tilde{J}_-(x) \; \Bigg\}  \label{subeq5:1}
\end{eqnarray}
In eq. 5a the renormalized velocities $\tilde{V}_\pm$, $V_\pm$ and
interaction parameters $K_\pm$ are given by:
\begin{eqnarray} \nonumber
V_+ = V_\uparrow a^2 + V_\downarrow b^2 + \frac{g_4}{\pi} a b \\
V_- = V_\downarrow a^2 + V_\uparrow b^2 - \frac{g_4}{\pi} a b
\label{subeq5:2}
\end{eqnarray}
The effect of the transformation in eq. 4a is to replace
$V_\uparrow$,$V_\downarrow$ by $V_\pm$.  The transformation in eq.
4b replaces $V_\pm$ by $\tilde{V}_\pm$, $\tilde{V}_\pm =
\frac{V_\pm}{(K_\pm + K_\pm^{-1})/2}$.  The interaction parameters
defined in eq. 4b take the values such that the first term in eq
3b has been eliminated.
\begin{equation}
K_\pm = \sqrt{\frac{1-R_\pm}{1+R_\pm}}, \quad R_\pm = \frac{g_2''
- g_1'' }{2 \pi V_\pm} \label{subeq5:3}
\end{equation}
Using $V_\uparrow = V_F + \frac{\Delta}{2}$, $V_\downarrow = V_F -
\frac{\Delta}{2}$, $\Delta > 0$ we observe from eq. 5c that for
$g_2'' - g_1'' > 0$, $K_+ > K_-$.  We will show tat due to the
anisotropy in the $K_\pm$ variables the wire will be polarized. In
eq. 5a we will choose the coefficient of $\tilde{J}_+(x)
\tilde{J}_-(-x)$ to vanish.  As a result we will determine the
parameters $a$ and $b$, $a^2 + b^2 = 1$.
\begin{equation}
a = \frac{1}{\sqrt{2}} \left[ 1 + \sqrt{\frac{1}{1+(g_2^\perp I +
g_4)/ \pi \Delta}} \right]^{1/2} \label{subeq5:4}
\end{equation}
where
\begin{equation}
  I = \frac{1 + K_-/K_+}{1 - K_-/K_+} \label{subeq5:5}
\end{equation}
\end{subequations}
\end{widetext}
For large $\Delta$ we introduce the parameter $\eta
\equiv \frac{g_2^+ I + g_4}{2 \pi \Delta} < 1$ and find $a \simeq
1 - \eta^2$, $b \simeq \eta^2$.  Once the coefficients $a$ and $b$
are determined eq.5a takes the form:
\begin{subequations} \label{eq6:whole}
\begin{equation}
H^{(1)} = \int_{-L}^L dx \Bigg\{ \pi \tilde{V}_+ \tilde{J}_+^2(x)
+ \pi \tilde{V}_- \tilde{J}_-^2(x) + 4 g_2^\perp (a^2 -
b^2)\frac{\sqrt{K_+ K_-}}{K_+ - K_-} \tilde{J}_+(x) \tilde{J}_-(x)
\Bigg\} \label{subeq6:1}
\end{equation}
The mixed term $\tilde{J}_+(x) \tilde{J}_-(x)$ is removed by the
linear transformation:
\begin{eqnarray} \nonumber
\tilde{J}_+(x) = \alpha \tilde{J}_\uparrow(x) + \beta
\tilde{J}_\downarrow(x) \\
\tilde{J}_-(x) = \beta \tilde{J}_\uparrow(x) + \alpha
\tilde{J}_\downarrow(x), \quad \alpha^2 + \beta^2 = 1
\label{subeq6:2}
\end{eqnarray}
We substitute eq. 6b into eq. 6a and obtain:
\begin{equation}
H^{(1)} = \int_{-L}^L dx \Bigg\{ \pi \tilde{V}_\uparrow
\tilde{J}_\uparrow^2(x) + \pi \tilde{V}_\downarrow
\tilde{J}_\downarrow^2(x) \Bigg\} \label{subeq6:3}
\end{equation}
where
\begin{eqnarray} \nonumber
\tilde{V}_\uparrow = \alpha^2 \tilde{V}_+ + \beta^2 \tilde{V}_- +
\frac{4 \alpha \beta}{\pi} g_2^\perp(a^2 - b^2)
\\
\tilde{V}_\downarrow = \alpha^2 \tilde{V}_- + \beta^2 \tilde{V}_+
- \frac{4 \alpha \beta}{\pi} g_2^\perp(a^2 - b^2) \label{subeq6:4}
\end{eqnarray}
The values of $\alpha$, $\beta$ are obtained by the requirement
that the coefficient of $ \tilde{J}_\uparrow(x)
\tilde{J}_\downarrow(x) $ vanishes.
\begin{equation}
 4 (a^2 - b^2)(\beta^2 - \alpha^2) g_2^\perp \frac{\sqrt{K_+ K_-}}{K_+ -
 K_-} + 2 \pi \alpha \beta (\tilde{V}_+ -\tilde{V}_-) = 0 \label{subeq6:5}
\end{equation}
with the solution,
\begin{equation}
\alpha = \frac{1}{\sqrt{2}} \big[ 1 + \frac{1}{\sqrt{1 +
\varepsilon^2}} \big]^{1/2}, \quad \varepsilon \equiv \frac{2
g_2^\perp (a + b^2)}{\pi \tilde{\Delta}} \frac{\sqrt{K_+ K_-}}{K_+
- K_-} \label{subeq6:6}
\end{equation}
\end{subequations}
where $\tilde{\Delta} = \tilde{V}_+ -\tilde{V}_-$.  Again for
$\varepsilon < 1 $ we obtain $ \alpha \approx 1 -
(\frac{\varepsilon}{2})^2$, $\beta \approx
(\frac{\varepsilon}{2})^2$.  The set of transformation 4a, 4b 6b
give the following relation between the ``free'' bosons
$\theta_\uparrow(x)$, $\theta_\downarrow(x)$ and the interacting
one $\tilde{\theta}_\uparrow(x)$, $\tilde{\theta}_\downarrow(x)$.
\begin{subequations} \label{eq7:whole}
\begin{eqnarray} \nonumber
\theta_\uparrow(x) = ( \tilde{\theta}_\uparrow(x) -
\tilde{\theta}_\uparrow(-x))( a\alpha K_+^{1/2} + b \beta
K_-^{1/2})\frac{1}{2} + ( \tilde{\theta}_\downarrow(x) -
\tilde{\theta}_\downarrow(-x))( a\beta K_+^{1/2} - b \alpha
K_-^{1/2})\frac{1}{2} \\
+ (\tilde{\theta}_\uparrow(x) + \tilde{\theta}_\uparrow(-x))(
a\alpha K_+^{-1/2} - b \beta K_-^{-1/2})\frac{1}{2} + (
\tilde{\theta}_\downarrow(x) + \tilde{\theta}_\downarrow(-x))(
a\beta K_+^{-1/2} - b \alpha K_-^{-1/2})\frac{1}{2}\;
\label{subeq7:1}
\end{eqnarray}
and similarly $\theta_\downarrow(x)$ is obtained from
$\theta_\uparrow(x)$ by replacing $K_+ \rightarrow K_-$, $K_-
\rightarrow K_+$.  In the limit that the ``flip'' interaction
parameters $g_4$ and $g_2^\perp$ are smaller than the Zeeman
interaction $\Delta$, we replace $a \sim 1 - \eta^2$, $b \sim
\eta^2$, $\alpha \simeq 1$, $\beta \simeq 0$ and find for $\eta <
1$,
\begin{eqnarray} \nonumber
\theta_\uparrow(x) \approx \frac{1}{2} (
\tilde{\theta}_\uparrow(x) - \tilde{\theta}_\uparrow(-x)) a
K_+^{1/2} + \frac{1}{2} ( \tilde{\theta}_\uparrow(x) +
\tilde{\theta}_\uparrow(-x)) a K_+^{-1/2} \\
- \frac{1}{2} (\tilde{\theta}_\downarrow(x) -
\tilde{\theta}_\downarrow(-x)) b K_-^{1/2} - \frac{1}{2}
(\tilde{\theta}_\downarrow(x) + \tilde{\theta}_\downarrow(-x)) b
K_-^{-1/2} \label{subeq7:2}
\end{eqnarray}
\end{subequations}
From eq. 7a we observe that the ``spin flip'' contribution is
reduced by a factor $b/a \sim \eta^2$.

Next we consider a metallic wire confined in the region $0 \leq x
\leq L$, for which we use again open boundary conditions.  We
bosonize the metallic hamiltonian using a periodic fermion field
$\Phi_\sigma(x) = \frac{1}{\sqrt{2\pi a}} \exp[i \sqrt{4 \pi}
\varphi_\sigma(x)]$, where $\varphi_\sigma(x)$ is a chiral boson.
The hamiltonian for the metal is given by,
\begin{equation}
H^{(2)} = \int_{-L}^L dx \Bigg\{ V_F(\partial_x
\varphi_\uparrow)^2 + V_F (\partial_x \varphi_\downarrow)^2
\Bigg\}
\end{equation}
In order to study tunnelling between the metal and the polarized
wire we consider the following tunnelling hamiltonian:
\begin{equation}
  H_T = t\Big[ \psi_\uparrow^+\big(-\frac{d}{2}\big)
  \Phi_\uparrow^+\big(\frac{d}{2}\big) + \psi_\downarrow^+\big(-\frac{d}{2}\big)
  \Phi_\downarrow^+\big(\frac{d}{2}\big) + h.c. \Big]
\end{equation}
Using eq. 7a we find for $H_T$ the representation:
\begin{eqnarray} \nonumber
H_T &\approx& \frac{t}{\pi d} \Big[ \cos\sqrt{4\pi}
\big(\tilde{\theta}_\uparrow(0) \gamma_{\uparrow\uparrow}^{1/2} +
\tilde{\theta}_\downarrow(0) \gamma_{\uparrow\downarrow}^{1/2} -
\varphi_\uparrow(0) \big) \\
&& \quad + \cos\sqrt{4\pi} \big(\tilde{\theta}_\downarrow(0)
\gamma_{\downarrow\downarrow}^{1/2} + \tilde{\theta}_\uparrow(0)
\gamma_{\downarrow\uparrow}^{1/2} - \varphi_\downarrow(0) \big)
\Big]; \quad d \ll L
\end{eqnarray}
In eq. 10 we have ignored the ``d'' dependence in the cosine term.
The parameters
$\gamma_{\uparrow\uparrow}$, $\gamma_{\downarrow\downarrow}$,
$\gamma_{\uparrow\downarrow}$, $\gamma_{\downarrow\uparrow}$ are
given by
\begin{eqnarray} \nonumber
\gamma_{\uparrow\uparrow}^{1/2} = a \alpha K_+^{- 1/2} + b \beta
K_-^{-1/2}, \quad \gamma_{\uparrow\downarrow}^{1/2}= a \beta
K_+^{- 1/2} - b \alpha
K_+^{1/2}, \\
\gamma_{\downarrow\downarrow}^{1/2} = b \beta K_+^{- 1/2} + a
\alpha K_-^{-1/2}, \quad \gamma_{\downarrow\uparrow}^{1/2}= b
\alpha K_+^{- 1/2} + a \beta K_-^{-1/2}.
\end{eqnarray}
We consider the hamiltonian $H = H^{(1)} + H^{(2)} + H_T$ in the
diagonal form given by $H^{(1)}$ in eq. 6c, and $H^{(2)}$ by eq. 8
and the tunnelling term $H_T$ given by eq. 10.  The effect of
tunnelling term $H_T$ is investigated within the Renormalization
Group (R.G.).  Performing a differential R.G. $\Lambda \rightarrow
\Lambda' = \Lambda - d\Lambda = \Lambda e^{-s}$, $s > 0$, we find
from the scaling dimensions of the field
$\tilde{\theta}_\uparrow$,  $\tilde{\theta}_\downarrow$,
$\varphi_\uparrow$,  $\varphi_\downarrow$, that the tunnelling
matrix elements $t^{\uparrow} = t$ and $t^{\downarrow} = t$ obey
the following scaling equations:
\begin{subequations} \label{eq10:whole}
\begin{eqnarray}
\frac{d t^{\uparrow}}{d s} = t^{\uparrow} \Big[1-\frac{1}{2}( 1 +
\gamma_{\uparrow,\uparrow} + \gamma_{\uparrow,\downarrow}) \Big]
\approx \frac{1}{2} t^{\uparrow} \Big[ 1 - \frac{a^2}{K_+} -
\frac{b^2}{K_-} \Big] \label{subeq10:1} \\
\frac{d t^{\downarrow}}{d s} = t^{\downarrow} \Big[1-\frac{1}{2}(
1 + \gamma_{\downarrow,\downarrow} + \gamma_{\downarrow,\uparrow})
\Big] \approx \frac{1}{2} t^{\downarrow} \Big[ 1 - \frac{a^2}{K_-}
- \frac{b^2}{K_+} \Big] \label{subeq10:2}
\end{eqnarray}
\end{subequations}
We solved eqs. 10a, 10b in terms of the length scale $e^s \equiv
\ell = \frac{1}{a} \min[L, L_T]$, we find:
\begin{subequations} \label{eq11:whole}
\begin{eqnarray}
G_\uparrow \simeq \frac{e^2}{h} |t|^2
\ell^{(\frac{a^2}{K_+}+\frac{b^2}{K_-}-1)} \label{subeq11:1}\\
G_\downarrow \simeq \frac{e^2}{h} |t|^2
\ell^{(\frac{a^2}{K_-}+\frac{b^2}{K_+}-1)} \label{subeq11:2}
\end{eqnarray}
\end{subequations}
The parameter$K_+$, $K_-$, $a^2$ and $b^2$ are given by equation
5c and 5d.  They obey the conditions $1 \geq K_+ \gg K_-$ and $a^2
\gg b^2$, $a^2 + b^2 = 1$.  We introduce the notations,
$\frac{a^2}{K_+}+\frac{b^2}{K_-}-1 \equiv
2(\frac{1}{K_\uparrow}-1)$, $\frac{a^2}{K_-}+\frac{b^2}{K_+}-1
\equiv2(\frac{1}{K_\downarrow}-1)$ and find the conductance
$G_\uparrow$ and $G_\downarrow$ in terms of the Luttinger
exponents, $K_\uparrow$, $K_\downarrow$.  From equation 5c we
observe that for repulsion interactions the anisotropy in the
velocities $V_\uparrow > V_\downarrow$ causes the anisotropy in
the $K_\pm$ values, $ 1 > K_+ > K_- > 0 $ and therefore
$1>K_\uparrow>K_\downarrow>0$.  For large Zeeman splitting $\Delta
= V_\uparrow - V_\downarrow$ we find $K_\uparrow \gg K_\downarrow$
and therefore a strong anisotropy in the conductance $G$ is
obtained, $G_\uparrow \gg G_\downarrow$.  The ratio between the
two conductances, $r \equiv G_\downarrow / G_\uparrow$
allows to define the polarization degree $P$.
\begin{subequations} \label{eq12:whole}
\begin{equation}
P = \frac{G_\uparrow-G_\downarrow}{G_\uparrow+G_\downarrow}
\approx \frac{1-r}{1+r} \label{subeq12:1}
\end{equation}
For long wires $L$ at temperature $T$ which obey $L>L_T$ the
infrared cutoff is provided by the thermal length and therefore we
find, $r \equiv (\frac{T}{T_F})^{2(\frac{K_\uparrow -
K_\downarrow }{K_\uparrow K_\downarrow})}$ where
\begin{equation}
2 \frac{K_\uparrow - K_\downarrow}{K_\uparrow  K_\downarrow} =
(\frac{g_2'' - g_1''}{2 \pi V_F})(\frac{\Delta}{V_F})(\frac{1}{1 -
(\frac{\Delta}{2 V_F})^2 }) > 0 \label{subeq12:2}
\end{equation}
\end{subequations}
From eq. 14 we observe that the polarization degree is enhanced at
low temperatures with the increase of the Zeeman term $\Delta$.

The interaction term $g_2'' - g_1''$ controls the polarization.
From eq. 3b we observe that the Luttinger parameter which controls
separately the spin up and spin down excitations is given by
$g_2'' - g_1''$.  For a Hubbard model (7) $g_2'' = g_1''$ and we
have no effect!  In order to obtain $g_2'' - g_1''\neq 0$ we need
an extended Hubbard model with large momentum transfer.  The
situation is similar to a Kondo problem where we have two Kondo
parameters - $J_\parallel$ and $J_\perp$.  For attractive
interactive interactions a spin gap is caused by a negative
$g_1^\perp$.  For $\Delta \neq 0$ the gap is avoided by the
oscillating term $2K_F (\frac{\Delta}{2 V_F})x$ which appears in
the ``cosine'' term (the last term in eq. 3b).  [More precisely we
can argue that $g_1$ becomes more negative under scaling.  The
growth of $|g_1|$ is stopped at the length scale $(exp
\ell^*)2K_F(\frac{\Delta}{2 V_F})d \simeq \pi$ (``d'' is the
lattice spacing).  Therefore for the length scale $\ell > \ell^*$
we can safely drop the ``cosine'' term in eq. 3b.]

In order to estimate the polarization at finite temperature
produced by the anisotropic tunnelling, $ K_\uparrow \neq
K_\downarrow$ we use the results given by eqs. 11, 12.  Using
typical values of $\frac{T}{T_F}\simeq 10^{-3} > \frac{a}{L}$,
$\frac{\Delta}{2 V_F} \simeq 0.5$ and $\frac{g_2'' - g_1''}{2 \pi
V_F } \simeq 0.5$, we obtain a polarization degree $P \simeq
80\%$.
Here we have investigated the case where the Zeeman splitting
is large, as a result we obtain two Luttinger liquids for spin up and
spin down.  In the opposite limit $ \Delta \rightarrow 0$ we obtain
two Luttinger liquids for spin and charge excitations.  The tunnelling
conductance will depend only on one exponent $K_c = K_\uparrow =
K_\downarrow$, the charge exponent (see ref. 5), but not separately on
$K_\uparrow$ and $K_\downarrow$.

To conclude the possibility for a magnetic wire and the simulation
of the F.M.-M.-F.M., transistor is suggested.  The idea is based
on the fact that the Luttinger interaction parameter $K_\uparrow
\neq K_\downarrow$.  As a result at the interface between two
systems one polarization will be reflected and only the second
polarization is able to tunnel into the second wire.

%----------------------------------------------------------------------

\end{document}